\documentclass[a4paper]{article}
\usepackage{qcircuit}
\usepackage{graphicx}
\usepackage{amsmath}
\usepackage[utf8x]{inputenc}
\usepackage[T1]{fontenc}
\usepackage{authblk}
\usepackage{subfig}
\usepackage[a4paper,top=3cm,bottom=2cm,left=3cm,right=3cm,marginparwidth=1.75cm]{geometry}
\input{amssym}
\usepackage[colorinlistoftodos]{todonotes}
\usepackage[colorlinks=true, allcolors=blue]{hyperref}
\usepackage{authblk}

\title{\bf Investigation of  Perturbation Theory  with Variational Quantum Algorithm}

\author[1]{H.Davoodi Yeganeh\thanks{h.yeganeh@tabrizu.ac.ir }}
\affil[1]{Faculty of Physics, Theoretical and astrophysics department , University of Tabriz, 51665-163 Tabriz, Iran}

\date{}

\begin{document}
\maketitle

\begin{abstract}

Variational Quantum Algorithms are among the most promising systems to implement quantum computing under the Noisy-Intermediate Scale Quantum (NISQ) technology. In  variational quantum
algorithm,  wavefunction represented by a parametrized  ansatz and variational parameters are
updated iteratively with a classical computer based on the measurement outputs from the quantum
computer. In this paper, we investigate  perturbation theory  with these algorithms and prospect the possibility of using the variational quantum algorithm to simulate quantum dynamics in perturbation theory. We illustrate the use of algorithms 
 with detailed examples which are in good agreement with analytical
calculations.

\end{abstract}
{\bf Keyword:} Variational  quantum algorithms, Perturbation Theory, Time-dependent Hamiltonian, Near-term devices, Quantum dynamics

\section{Introduction}

We know that efficiently simulating quantum systems with high degrees of freedom  is hard for classical computers due to the exponential growth of variables for characterizing these systems \cite{Feynman}. Quantum computers were proposed to solve such an exponential explosion problem, ranging from optimization to materials design, and the algorithms used in quantum computers have made great strides in the calculation and efficiency of various issues \cite{Aram,Otterbach,Peruzzo,Farhi}. Among the different approaches to quantum computing, the near-term quantum devices are mostly center around quantum simulations, which consists of a relatively low-depth quantum circuit by variational quantum algorithms. The variational algorithms were recently attracting a lot of attention, designed to utilize both quantum and classical resources to solve specific optimization tasks not accessible to traditional classical computers \cite{Wecker,McClean,r8,aa1,aa2}. The main idea of this method is to divide the problem into two parts that each of performing a single task and can be implemented efficiently on a classical and a quantum computer. The significant benefit of this method is that it gives rise to a setup that can have much less strict hardware requirements and promising for NISQ \cite{Preskill} and devices typically have on the order of  fewer qubits(contain from $ 10$ to $ 10^3$ of qubits) with high gate fidelity and not fault-tolerant error correction.
From a practical point of view, most of the current efforts concentrate on analog quantum computing methods such as quantum annealing \cite{Kadowaki,Finnila}, and quantum adiabatic simulation\cite{Goldstone,Babbush}.
Recently, several variational quantum algorithms for specific tasks have been developed, and analog approaches can be approximately solved using gate model NISQ devices. These algorithms and their applications are progressing in various fields such as  variations quantum eigensolver (VQE) which is a hybrid algorithm to approximate the ground state eigenvalues for quantum simulations \cite{Kandala,Peruzzo}, quantum approximate optimization algorithm (QAOA) for finding an approximate solution of an optimization problem\cite{Farhi,Farhi2}, dissipative-system Variational Quantum Eigensolver(dVQE) to simulate Non-equilibrium steady states an open system \cite{Yoshioka}, molecular simulations on a quantum computer \cite{Grimsley},variational quantum state diagonalization (VQSD) \cite{LaRose}  , also to simulate dynamics of open and closed systems\cite{H1,H2}. A complete overview of these algorithms is provided in Ref.\cite{RR2021,jap1}.
 In quantum mechanics, few problems have exact solution, whether or not Hamiltonian is time-dependent. Hence, approximate methods have been proposed\cite{sa}. The approximate methods play a significant role in the study of different quantum systems. In general, we have three approximate methods; they are Perturbation Theory (PT)\cite{k21}, Variational theory\cite{vv0} and WKB approximation\cite{wkb}.
Here we focuse on  PT, suppose there is a problem the Hamiltonian of which is represented by $H$, for which the exact solution is not known. The Hamiltonian of the system can be described in terms of two terms; the Hamiltonian is written $H=H_0+\lambda V$, in which $H_0$ is called as the unperturbed Hamiltonian, and  $V$ is called as the perturbation Hamiltonian, and $\lambda$ is a continuous real parameter. It is important to realize that the system actually can be split in terms of these two Hamiltonians, and that should be known. In addition, there are two more conditions; one is that solution of the  unperturbed Hamiltonian  is completely known. The following condition, which is a stringen condition, that $\lambda$ has to be much smaller than 1 $(\lambda <<1)$, which means the perturbation terms are much smaller compared to the original Hamiltonian $H_0$. In general, the PT can be divided into two categories, Time-Independent Perturbation Theory(TIPT) and Tim-Dependent Perturbation Theory(TDPT). 
In TIPT  perturbation Hamiltonian, $V$ is independent of time, and in TDPT $V$ is dependent of time. 
Here, we consider TIPT, and TDPT, and investigate them with the variational quantum algorithm. In fact, in TDPT, we are dealing with a time-dependent Hamiltonian and  try to find its spectrum (In the language of mathematics  described in detail in Section 2). In this paper, we prospect the possibility of using a variational quantum algorithm to simulate
quantum dynamics in TDPT and, TIPT. For this purpose, we employ time-dependent  variational quantum algorithm introduced in
Ref\cite{H1,H2,Li2016EfficientMinimisation,e14,chen}. In this algorithm, the time-dependent quantum state is approximated
by a parametrized quantum state and using McLachlan’s principle, the equation of motion for the
variational parameters is obtained(described in detail in Section 3). The paper is organized as follows: In Section 2, we  introduce Time-Independent and  Time-Dependent Perturbation Theory. The variational quantum algorithm for quantum dynamics is described in Section 3. In Section 4, we present the numerical results. Finally, Section 5, gives the conclusions.

\section{ Perturbation Theory}
Perturbation theory is widely used and plays an important role in describing real quantum systems, because it is impossible to find
exact solutions to the Schrodinger equation for Hamiltonians
even with moderate complexity. In general, the PT can be divided into two categories, TIPT and TDPT. In this section, we describe these  two categories in detail.
\subsection{Time-Independent Perturbation Theory}
Time-independent perturbation theory is a mathematical tool for investigating quantum system which Hamiltonian is  independent of time. Consider the time- independent Schrodinger equation
\begin{equation}
H|\psi_n\rangle =(H_0+\lambda V)|\psi_n\rangle=E_n|\psi_n\rangle,
\end{equation}
where $|\psi_n\rangle$ and $E_n$ are the  $n^{th}$ eigenstate and energy respectively.  The $H_0$ is unperturbed Hamiltonian and satisfies the time independent Schrodinger equation.
 Since,
in TIPT,  eigenstate and energy  will be sought in the form of an expansion in powers of $\lambda$,
After som calculation,  in the first order approximation  eigenstates and energies are expressed as

$$E_n=E_n^{(0)}+\langle n^{(0)}| V|n^{(0)}\rangle$$
\begin{equation}\label{E01}
|\psi_n\rangle=|n^{(0)}\rangle -\sum _{m\neq n} \frac { V_{mn}}{E_m^{(0)}-E_n^{(0)}}|n\rangle.
\end{equation}
For second-order approximation, we have
$$E_n=E_n^{(0)} + \sum _{m\neq n} \frac {|V_{mn}|^2}{E_m^{(0)}-E_n^{(0)}},$$
\begin{equation}\label{E03}
|\psi_n\rangle=|n^{(0)}\rangle -\sum _{m\neq n} \frac { V_{mn}}{E_m^{(0)}-E_n^{(0)}}|m\rangle -\frac{1}{2}\sum _{m\neq n} \frac { |V_{mn}|^2}{E_m^{(0)}-E_n^{(0)}}|n\rangle +\sum _{m\neq n} [\sum _{l\neq n} \frac{V_{mn}V_{ln}}{(E_m^{(0)}-E_n^{(0)})(E_l^{(0)}-E_n^{(0)})}]|m\rangle.
\end{equation}
Denote the
$V_{mn}$ and $V_{ln}$ are expectation value of $V$ as $V_{m(l)n}=\langle m(l)|V|n\rangle$.

\subsection{Time-Dependent Perturbation Theory}

In real world,  there are many quantum systems
of importance with time dependence, therefore, the dynamics of these systems must be examined by considering the time dependence. For a system described by a Hamiltonian $H_0$, which is time–independent, the most general state of the system can be described by a wavefunction $|\psi(t)\rangle$ which can be expanded in the energy eigenbasis $\{|n\rangle\} $as follows
\begin{equation}
|\psi(t)\rangle = \sum_n c_n \text{ exp}(−iE_nt/\hbar )|n\rangle,
\end{equation}
where the coefficients,$c_n$ are time-independent and $E_n$ is the eigenvalue corresponding to the energy eigenstate $|n\rangle$ of $H_0$. For time dependent case, we consider a Hamiltonian $H$ as it can be split into two parts, that's mean Hamiltonian is of the form
\begin{equation}
H=H_0+V(t),
\end{equation}
we can again expand in, $|n\rangle$  the time-independent eigenbasis of $H_0$
\begin{equation}
|\psi(t)\rangle = \sum_n c_n(t) \text{ exp}(−iE_nt/\hbar )|n\rangle,
\end{equation}
but the coefficients, $c_n$ will now in general be time-dependent. By using interaction picture and writing $E_n=\hbar \omega_n$ and given that 
the wavefunction satisfies the time-dependent Schrödinger equation, we obtain

$$\sum_n (i\hbar \dot{c}_n − c_n{V} ) \text{ exp}(−i\omega_n t)|n \rangle = 0. $$
With simple calculations we have
$$i\hbar \dot{c}_m \text{ exp}(−i\omega_m t) − \sum_n c_n V_{mn} \text{ exp}(−i\omega_nt) = 0 ,$$
giving the following set of coupled, first–order differential equations for the coefficients
\begin{equation}\label{E02}
i\hbar \dot{c}_m = \sum_n c_nV_{mn} \text{ exp}(i\omega_{mn}t) ,
\end{equation}
where $\omega_{mn}=\omega_n-\omega_n$ and $V_{mn}=\langle m|V|n\rangle$
It can be seen that,  the coefficient $c_m$ varies with time  i.e. the probability that a measurement will show the system to be in the $m^{th}$ eigenstate. It is exact, but not terribly useful because we must, in general, solve an infinite set of coupled differential equations.
Here, instead of solving differential equations, we obtain the dynamics of the time-dependent Hamiltonian  with the variational quantum algorithm directly.

\section{Variational Quantum Algorithm}
For simulate the time-inpendent Hamiltonian dynamics, we employ variational quantum algorithm introduced in
Ref\cite{H1,H2,Li2016EfficientMinimisation,e14,chen}.
 In this algorithm, state  $|\psi(t)\rangle$ is approximated
by a parametrized  state $|\phi(\vec{\lambda})\rangle$, i.e.
$$|\psi(t)\rangle= exp(-i Ht) |\psi(0)\rangle \approx |\phi(\vec{\lambda})\rangle,$$
where $\vec{\lambda}=\{\lambda_1(t),\lambda_2(t)....\}$ are variational parameters. By using time-dependent variational principle corresponding
to the Schrödinger equation we have
\begin{equation}
\mathcal{L}=\langle \psi(t)|\frac{i\partial}{\partial t}-H|\psi(t) \rangle
\end{equation}
where, $\mathcal{L}$ is
Lagrangian. By considering $ |\psi(t)\rangle \approx |\phi(\vec{\lambda})\rangle$ and using the Euler-Lagrange equation we have 
\begin{equation}
\sum_q M_{pq}\dot{\lambda_q}=V_p,
\end{equation}
where

\begin{equation}
M_{p,q}=i \frac{\partial \langle \phi(\vec{\lambda})|}{\partial \lambda_k} \frac{\partial |\phi(\vec{\lambda})\rangle}{\partial \lambda_q} + H.C
\end{equation}
\begin{equation}
V_{p}= \frac{\partial \langle \phi|}{\partial \lambda_p}H |\phi(\vec{\lambda})\rangle  + H.C
\end{equation}

For simulate the time-dependent Hamiltonian dynamics, we employ variational quantum algorithm introduced in
Ref\cite{H1,H2,Li2016EfficientMinimisation,e14,chen}. In this algorithm, state  $|\psi(t)\rangle$ is approximated
by a parametrized  state $|\phi(\vec{\lambda})\rangle$, i.e.
$$|\psi(t)\rangle= T exp(-i\int^t _0  H(t')dt')|\psi(0)\rangle \approx |\phi(\vec{\lambda})\rangle,$$
where $\vec{\lambda}=\{\lambda_1(t),\lambda_2(t)....\}$ are variational parameters and $T$ is the time-ordering operator. By using  McLachlan’s principle,( or Dirac and Frenkel variational principle) the equation of motion for the
variational parameters is obtained by minimizing the quantity $||(i\frac{\partial}{\partial t}-H)|\phi(\vec{\lambda})\rangle||$, so we have
$$\lambda(t_{n+1})=\lambda(t_n)+\dot{\lambda} \delta t$$
\begin{equation}\label{eul}
\sum_i M_{ki} \dot{\lambda_i}=V_k.
\end{equation}
where 
\begin{equation}\label{eul}
\begin{split}
 M_{ki}=Re\langle \frac{\partial \phi(\vec{\lambda})}{\partial \lambda_k}|\frac{\partial \phi(\vec{\lambda})}{\partial \lambda_i}\rangle \\  V_k=Im\langle \phi(\vec{\lambda})|H|\frac{\partial \phi(\vec{\lambda})}{\partial \lambda_k}\rangle.
\end{split}
\end{equation}
The coefficients of the differential Eq.( \ref{eul}) are determined
using a quantum computer, while each propagation step is carried out by classically
solving the differential equation.
To obtain high accuracy of  algorithm, we must be sensitive in choosing the wavefunction ansatz. Here we consider an ansatz of the form
$$|\phi(\vec{\lambda})\rangle=U(\vec{\lambda})|\phi_0\rangle=U(\lambda_1)U(\lambda_2)...U(\lambda_i)...U(\lambda_N)|\phi_0\rangle$$
The evolution operator is unitary, so it is equivalent to a certain rotation in the Hilbert
space of states .i.e.
\begin{equation}\label{unit22}
U(\lambda_i)=exp(-i \lambda_i \Lambda_i)=exp(-i\sum_{j} \lambda_i s_{i,j} \hat{\sigma}_{i,j}),
\end{equation}
where $\Lambda_i=\sum_{j} s_{i,j} \hat{\sigma}_{i,j}$ where $\hat{\sigma}_{i,j}$ are Pauli operators.
So, we have,
\begin{equation}\label{unider1}
\begin{split}
\frac{\partial U(\vec{\lambda})}{\partial \lambda_i}=U(\lambda_1)U(\lambda_2)...-i \sum_{j} s_{i,j} \hat{\sigma}_{i,j} U(\lambda_i)....U(\lambda_N)\\
\frac{\partial U^\dagger(\vec{\lambda})}{\partial \lambda_i}= U^\dagger(\lambda_N)....i \sum_{j} s^*_{i,j} \hat{\sigma}_{i,j} U^\dagger(\lambda_i)...U^\dagger(\lambda_2)U^\dagger(\lambda_1),
\end{split}
\end{equation}
therefore the matrix elements of $M$ and $V$ are
\begin{equation}\label{unider1}
\begin{split}
M_{ki}=Re(\langle \phi_0|U(\lambda_1)^\dagger...U(\lambda_N)^\dagger \Lambda_k^\dagger...U(\lambda_l)^\dagger \Lambda_iU(\lambda_i)....U(\lambda_1)|\phi_0\rangle)\\
V_k=Im(i\sum_j c_j\langle \phi_0|U(\lambda_1)^\dagger...U(\lambda_N)^\dagger h_j U(\lambda_N)...\Lambda_k U(\lambda_k)... U(\lambda_1)|\phi_0\rangle),
\end{split}
\end{equation}
where the Hamiltonian expressed $H=\sum_j c_j h_j$. The $M_{ki}$ and $V_k$ are
obtained in a quantum circuit via the Hadamard or Swap test. Note that the matrix elements in the first and second-order approximations i.e. Eq(\ref{E01}), Eq(\ref{E03}) and Eq(\ref{E02})   can be obtained by  variational quantum algorithm.

\section{Numerical Example }

\begin{figure}[h!]
\centering
\includegraphics[width=9cm]{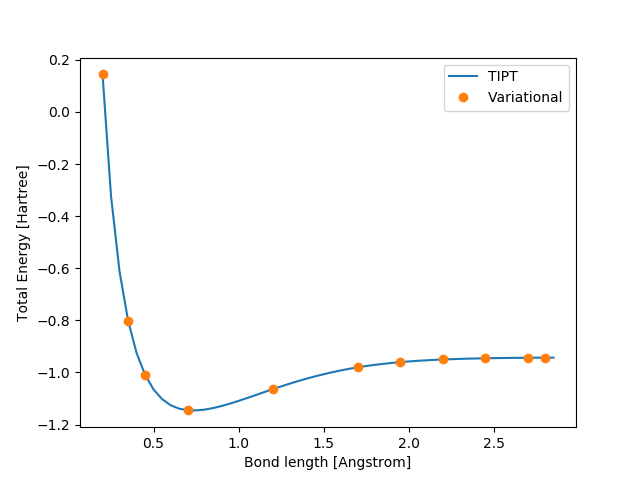}
\caption{Energy curves of the molecular hydrogen ground state in the two cases TIPT and Variational method }\label{fig66}
\end{figure}

\begin{figure}[h!]
\centering
\includegraphics[width=11cm]{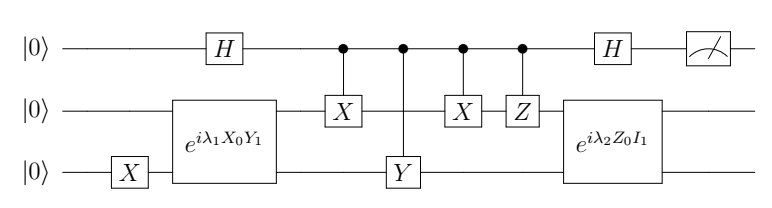}
\caption{Quantum circuit used in the finding ground
state energy  $H_2$ molecule. This  circuit employed to   evaluate coefficients. The ancilla qubit on the top
line undergoes Hadamard gates $H$ and unitary operations nd controls
operations apply on initial state $|01\rangle$  }\label{fig666}
\end{figure}

In this section, we numerically test the performance of the previously described  variational quantum algorithm on some of the systems. In the first example, consider $H_2$ molecule and find ground
state energy  $H_2$. We used a standard
molecular basis set, the minimal STO-3G basis. Via JordanWigner or Bravyi-Kitaev transformation, the qubit-Hamiltonians of this molecule can be obtained.
So, 
\begin{equation}
H=g_0 I+g_1 Z_0+g_2 Z_1+g_3 Z_0 Z_1+g_4 Y_0 Y_1+g_5 X_0 X_1,
\end{equation}
where the coefficients $g_i$ were all derived in this classical
preprocessing step and 
$\{X_i,Y_i,Z_i\}$ are Puali matrixs.
Whit using TIPT we can divid this Hamiltinain in two trem,
$$H_0=g_0 I+g_1 Z_0+g_2 Z_1+g_3 Z_0 Z_1$$
$$H'=g_4 Y_0 Y_1+g_5 X_0 X_1$$
$H_0$ is the unperturbed Hamiltonian and  $H'$ is  the perturbation Hamiltonain. $H_0$ is diagonal matrix and we can obtain eigenstate and energy $H$ by using  TIPT.  For  variational quantum algorithm, we consider an anstaz of the form 
$$|\psi(\vec{\lambda})\rangle=e^{i\lambda_1 X_0Y_1}e^{i\lambda_2 Z_0I_1}|01\rangle$$. Output of variational algorithm compared it to TIPT solution
ground
state energy as shown in figure \ref{fig66}. Also we implement a quantum circuitfor evaluating coefficients on
quantum processor in figure \ref{fig666}

\begin{figure}[h!]
\centering
\includegraphics[width=11cm]{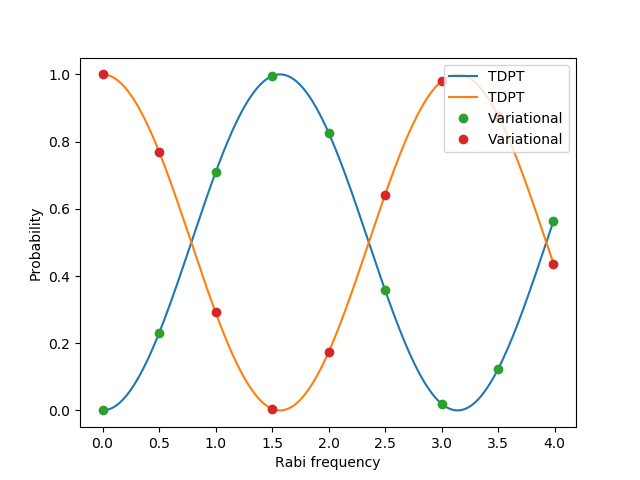}
\caption{Probability of transition in the two cases TDPT and Variational method for $t=1$. The orange and blue lines  corresponds to the probability $|c_1(t)|^2$ and $|c_2(t)|^2$ respectively. }\label{fig56}
\end{figure}

\begin{figure}[h!]
\centering
\includegraphics[width=11cm]{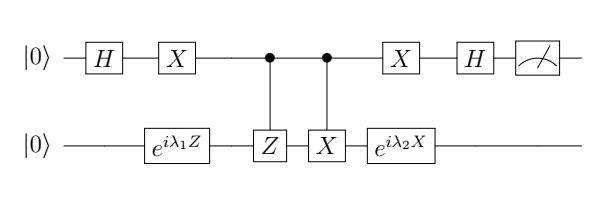}
\caption{Quantum circuit used in the second example to   evaluate coefficients. The ancilla qubit on the top
line undergoes Hadamard gates $H$ and unitary operations nd controls
operations apply on initial state $|0\rangle$  }\label{fig25}
\end{figure}

In  second example let us consider a
two-state system with
$$H_0=E_1 |0\rangle \langle 0| + E_2 |1\rangle \langle 1|$$
$$V(t)=\delta e^{i\omega t}|0\rangle \langle 1| + \delta e^{-i\omega t}|1\rangle \langle 0|.$$
Whit TDPT, we can obtain
$$i\hbar \vec{\dot{c}} = \delta(e^{i(\omega -\omega_{21})t}|0\rangle \langle 1| +e^{-i(\omega -\omega_{21})t}|1\rangle \langle 0|) \vec{{c}},$$
where $\vec{{c}}$ is the two-component vector $\vec{{c}}=(c_1(t),\ c_2(t)) $and $\omega_{21}=(E_2-E_1)/\hbar$. With the initial condition $ c_1(0)=1$, and 
$c_2(0)=0$ this
equation has the solution
\begin{equation}
|c_2(t)|^2=\frac{4\delta ^2}{\delta^2 +\hbar^2 (\omega-\omega_{21})^2}sin^2\Omega t \qquad |c_1(t)|^2=1-|c_2(t)|^2
\end{equation}
where $\Omega=((\delta/\hbar)^2+(\omega-\omega_{21})^2/4)^{1/2}$ is known as the Rabi frequency. The solution,
which varies periodically in time, describes the transfer of probability from state 0 to
state 1 and back.
For  variational quantum algorithm, we consider an anstaz of the form
$$|\psi(\vec{\lambda})\rangle=e^{i\lambda_1 Z}e^{i\lambda_2 X}|0\rangle$$
Output of variational algorithm compared it to TDPT solution the transfer of probability from states  as shown in figure \ref{fig56}. Also  We propose a quantum circuit for evaluating coefficients on quantum processor in figure \ref{fig25}

\newpage

\section{Conclusion}
 We have considered a variational quantum algorithm for investigation of perturbation theory.
We considered perturbation theory as two categories, time-independent perturbation theory  and time-dependent perturbation theory.
 We have used   a variational quantum algorithm which, exact state   is approximated
by a parametrized  state and variational principle. In the case TIPT and, TDPT
 using the time-dependent variational principle  and McLachlan’s principle corresponding
to the Schrödinger equation, respectively, we obtained the differential equations for simulating dynamics.
We compare the analytical and numerical results obtained from the exact solution with the variational quantum algorithm which, are in good
agreement with each other. 

\bibliographystyle{unsrt}
\bibliography{vosq}


\end{document}